\documentclass[11pt]{article}
\usepackage[a4paper,top=80pt,lmargin=80pt,rmargin=80pt,bottom=80pt]{geometry}

\usepackage{graphicx}

\usepackage[]{cite}

\expandafter\let\csname equation*\endcsname\relax
\expandafter\let\csname endequation*\endcsname\relax

\usepackage{amsmath}
\usepackage[]{amssymb}
\usepackage[]{mathrsfs}
\usepackage[]{eucal}
\usepackage[]{tensor}
\usepackage[]{amsthm}
\usepackage[]{bm}

\usepackage{tikz}
\usepackage{tikz-cd}
\newcommand{\btkz}{\begin{tikzpicture}}
\newcommand{\etkz}{\end{tikzpicture}}

\newcommand{\pf}[1]{\mathbf{#1}}
\newcommand{\dd}{\partial}
\newcommand{\hdg}{\star} 
\newcommand{\df}{\mathrm{d}}
\newcommand{\w}{\wedge}
\newcommand{\veps}{\bm{\epsilon}}
\newcommand{\Lie}{\pounds}
\newcommand{\nab}[1]{\nabla_{\!#1}}
\newcommand{\qqd}{\ , \quad}
\newcommand{\bc}{\begin{center}}
\newcommand{\ec}{\end{center}}
\newcommand{\be}{\begin{equation}}
\newcommand{\ee}{\end{equation}}

\newcommand{\F}{\pf{F}}
\newcommand{\Z}{\pf{Z}}
\newcommand{\A}{\pf{A}}

\newcommand{\FF}{\mathcal{F}}
\newcommand{\GG}{\mathcal{G}}

\newcommand{\LL}{\mathscr{L}}

\newcommand{\defeq}{\mathrel{\mathop:}=}

\usepackage{dsfont}

\newcommand{\rr}{\mathds{R}}

\usepackage[unicode]{hyperref}
\usepackage{xcolor}
\definecolor{pastgreen}{HTML}{669900}
\definecolor{pastblue}{HTML}{336699}
\definecolor{pastred}{HTML}{990000}
\definecolor{linkcol}{HTML}{663333}
\hypersetup{colorlinks,linkcolor={pastblue},citecolor={pastblue},urlcolor={pastblue}}

\theoremstyle{plain} \newtheorem{tm}{Theorem}[section]
\theoremstyle{plain} \newtheorem{lm}[tm]{Lemma}
\theoremstyle{plain} \newtheorem{defn}[tm]{Definition}
\newcommand{\btm}{\begin{tm}}
\newcommand{\etm}{\end{tm}}
\newcommand{\blm}{\begin{lm}}
\newcommand{\elm}{\end{lm}}
\newcommand{\bdefn}{\begin{defn}}
\newcommand{\edefn}{\end{defn}}

\begin{document}

\begin{flushright}
\texttt{ZTF-EP-26-07}

\texttt{RBI-ThPhys-2026-18}
\end{flushright}

\vspace{20pt}

\bc
{\huge Constraints on regular black holes with nonminimally coupled electromagnetic fields}

\vspace{25pt}

{\Large Ana Bokuli\'c$^a$, Tajron Juri\'c$^b$, Luka Kanai Peji\'c$^a$, 

Filip Po\v zar$^b$ and Ivica Smoli\'c$^a$}

\bigskip

e-mails: abokulic@phy.hr, tjuric@irb.hr, luka.kanai.pejic@student.pmf.hr, fpozar@irb.hr, ismolic@phy.hr

\medskip

$^a$Department of Physics, Faculty of Science, University of Zagreb, Bijeni\v cka cesta 32, 10000 Zagreb, Croatia

\smallskip

$^b$Rudjer Bo\v skovi\'c Institute, Bijeni\v cka cesta 54, HR-10002 Zagreb, Croatia
\ec

\vspace{15pt}

\begin{abstract}
Construction of physically realistic theories admitting regular black hole solutions remains an important open problem in gravitational physics. While theories with electromagnetic fields minimally coupled to gravity have been extensively studied over the past two decades, theories with nonminimal couplings remain comparatively unexplored. We investigate theories containing the interaction terms $R F_{ab}F^{ab}$, $R_{ab} \tensor{F}{^a_c} F^{bc}$ and $R_{abcd} F^{ab} F^{cd}$, which generically arise in low-energy effective Lagrangians. We prove that magnetically charged regular black holes are excluded, except possibly for a special choice of coupling constants, and argue that a similar conclusion applies to electrically charged regular black holes. We further show that similar conclusions hold for Lagrangian terms of the form $f(R,F_{ab}{\star F}^{ab})$.
\end{abstract} 

\vspace{5pt}


\bigskip

\section{Introduction} 

The Einstein--Maxwell theory provides a well-established classical description of electrodynamics in curved spacetime, yet its corresponding black hole solutions generically contain curvature singularities. This limitation motivates considering modifications of the gravitational and electromagnetic sectors, beyond the simplest, minimally coupled Einstein--Maxwell action. One possible direction is the inclusion of the direct couplings between the electromagnetic field and spacetime curvature. From the effective field theory (EFT) perspective, such corrections are expected as the minimally coupled action corresponds only to the leading term in the low-energy expansion. Additional nonminimal couplings naturally arise as higher-dimensional operators, suppressed by the cutoff scale and constrained by diffeomorphism and gauge symmetries. Even so, these principles allow an infinite tower of terms\footnote{An exhaustive classification of scalars formed by contracting the electromagnetic field tensors with Riemann and Ricci tensors, along with the equations of motion for a broad class of nonminimally coupled theories, can be found in \cite{BL05}.} and the resulting theories might exhibit potentially problematic features, such as ghost degrees of freedom and instabilities. To avoid some of these issues, one may demand that the equations of motion remain second order. At the classical level, imposing this requirement, Horndeski derived a unique vector-tensor action \cite{HOR76}. Interestingly, Horndeski vector-tensor (HVT) theory can also be obtained via dimensional reduction of the Gauss--Bonnet theory \cite{Buch79, MMHS88_1}, suggesting that nonminimally coupled terms could also emerge from higher-dimensional theories. 

\smallskip

Black hole solutions in nonminimally coupled theories may reveal qualitatively distinct properties compared to the Reissner--Nordstr\"om solution, as illustrated in the early numerical analysis using HVT as a model \cite{MHS88}. Later, charged black hole solutions were obtained within the three-parameter nonminimal theory studied in \cite{BBL08, BZ15}. The number of parameters is further reduced to a single one by imposing additional physical requirements, such as the regularity of the electric field. The remaining coupling constant is proposed to be, with a degree of ambiguity, related to the effective radius at which nonminimal interactions become dominant \cite{BZ15}. Alternatively, the value of coupling parameters can be matched with QED: vacuum polarization effects on curved spacetime fix them in terms of the fine structure constant \cite{DH80}. 

\smallskip
Besides altering black hole structure, nonminimal couplings also lead to certain phenomenologically relevant effects. In general, nonminimal extensions of Einstein--Maxwell system break its conformal invariance and the SO(2) duality symmetry \cite{CM21, CM21_3}. The loss of conformal invariance may have important cosmological implications, for example, it could provide a mechanism for primordial magnetogenesis \cite{TW88}. Nonminimal interactions can significantly affect magnetic fields in the vicinity of black hole horizons, opening the window for astrophysical tests capable of constraining the coupling strength \cite{PS19, RPS23}. Furthermore, the modified propagation of light in these theories can produce distinct lensing signatures and shadow images \cite{CRDEF25, YGZ26, Vagnozzi:2022moj}.

\smallskip

On the other hand, the standard Einstein--Maxwell theory may be modified by replacing Maxwell’s electrodynamics with its nonlinear generalizations, resulting in a broad family of nonlinear electrodynamics (NLE) theories. These models, originating in the 1930s, either from the effective quantum field theory \cite{HE36} or phenomenological considerations \cite{Born34, BI34}, also found applications in the context of gravitational theory. The interpretation of the regular Bardeen black hole \cite{Bardeen68} within the NLE framework \cite{ABG00} suggested that NLE fields could tame black hole curvature singularities. However, the failure of this model to reduce to Maxwell’s theory in the weak field limit pointed to the possible shortcomings of this proposal. Subsequent analyses revealed fundamental obstructions; electrically charged regular black holes are incompatible with Maxwell’s weak-field behavior of the NLE source, while attempts to regularize the dyonic and magnetic configurations also face limitations \cite{Bronnikov00, BSJ22}. Even if one circumvents these no-go theorems, the black hole mass and charge have to be fine-tuned in the regularized solutions \cite{BJS26}. Furthermore, NLE theories leading to regular black holes have been found to violate causality\footnote{This is consistent with the fact that causality implies certain energy conditions whose validity excludes regular solutions via singularity theorems \cite{Russo1}.}\cite{Russo2}. Under a given set of assumptions, similar constraints can be found in the HVT-NLE theory: it does not admit regular magnetically charged black holes, while in the electrically charged case, a regular branch may be present \cite{CFTS26}.

\smallskip

Compared to the NLE case, black hole regularization within higher-derivative generalizations of Einstein--Maxwell theory might offer several advantages. In \cite{CM21_2}, it has been shown that a particular non-minimal theory admits a regular black hole solution with arbitrary mass and charge parameters, thereby avoiding the mass-charge relation. However, another regular solution can be interpreted either as arising from an NLE model \cite{BALART201414} or a nonminimally coupled theory \cite{Sert16}, requiring fine-tuning of the mass and charge parameters in both cases. Since the black hole properties strongly depend on the type of nonminimal coupling, it is not straightforward to determine when regular solutions can be expected or how general this possibility is. We address this challenge by providing a systematic analysis that rules out certain theories as candidates for regularization. 

\smallskip

The paper is organized as follows. In Section 2 we summarize basic elements (Lagrangian terms, form of the metric and the electromagnetic field, regularity assumptions) that will be used in later analysis. In Section 3 we investigate nonminimal coupling in which Ricci scalar is multiplied with some of the electromagnetic invariants, while in Section 4 and Section 5 we proceed with the Lagrangian terms of the form, respectively, $R_{ab} \tensor{F}{^a_c} F^{bc}$ and $R_{abcd} F^{ab} F^{cd}$. In Section 6 we discuss implications of the result obtained in the paper and further open questions. In the Appendix A we collect most important field variations and in the Appendix B we prove that, under the assumed symmetries, the electric and magnetic 1-forms are purely radial.

\smallskip

\emph{Notation and conventions}. We use the ``mostly plus'' metric signature and natural system of units in which $c = G = 4\pi\epsilon_0 = 1$. The two elementary electromagnetic invariants are denoted by $\FF \defeq F_{ab} F^{ab}$ and $\GG \defeq F_{ab} {\hdg F}^{ab}$, where the Hodge dual is ${\hdg F}_{ab} \defeq \tensor{\epsilon}{_a_b^c^d} F_{cd}/2$. Einstein tensor $G_{ab}$ and the Maxwell's electromagnetic energy-momentum tensor $T^{\mathrm{(Max)}}_{ab}$ are defined as follows
\begin{align}
G_{ab} & \defeq R_{ab} - \frac{1}{2} \, R g_{ab} , \\
T^{\mathrm{(Max)}}_{ab} & \defeq \frac{1}{4\pi} \left( F_{ac} \tensor{F}{_b^c} - \frac{1}{4} \, \FF g_{ab} \right) .
\end{align}
From the menagerie of Bachmann--Landau symbols we recall that
\begin{itemize}
\item[(a)] $f = O(g)$ as $r \to 0^+$ if there are constants $C,\delta > 0$ such that $|f(r)| \le C |g(r)|$ for all $0 < r < \delta$;
\item[(b)] $f = o(g)$ as $r \to 0^+$ if for any $\epsilon > 0$ there exists $\delta(\epsilon) > 0$ such that $|f(r)| \le \epsilon |g(r)|$ for all $0 < r < \delta(\epsilon)$; and 
\item[(c)] $f = \Omega(g)$ as $r \to 0^+$ if there are constants $C,\delta > 0$ such that $|f(r)| \ge C |g(r)|$ for all $0 < r < \delta$. 
\end{itemize}
For example, assuming that we are looking at $r \to 0^+$ limit, $f = O(r^{-n})$ implies that the function $f$ grows as most as $r^{-n}$, $f = o(r^{-n})$ implies that the function $f$ grows strictly slower than $r^{-n}$, while $f = \Omega(r^{-n})$ implies that the function $f$ does not grow slower than $r^{-n}$.

\section{Basic setup} 

We assume that the action consists of the standard Einstein--Maxwell term and a nonminimal coupling between the electromagnetic and gravitational fields,
\be
S = \frac{1}{16\pi} \int \big( R - 2\Lambda - \FF + \ell(g_{ab}, \epsilon_{abcd}, R_{abcd}, F_{ab}) \big) \, \veps .
\ee
In general, the nonminimal coupling term $\ell$ is a function of scalars obtained by contractions of the spacetime metric $g_{ab}$, the Levi--Civita tensor $\epsilon_{abcd}$, the Riemann tensor $R_{abcd}$, and the electromagnetic tensor $F_{ab}$. In this paper we focus on the simplest families of nonminimal coupling, which we refer to as a type A terms if $\ell = f_A(R,\FF,\GG)$, type B terms if $\ell = f_B(R_{ab} \tensor{F}{^a_c} F^{bc})$ and type C terms if $\ell = f_C(R_{abcd} F^{ab} F^{cd})$. Key variations, used in derivations of the field equations, are gathered in the Appendix \ref{app:Var}.

\smallskip

We assume that the spacetime is static, with the corresponding Killing vector field $k^a$, and spherically symmetric, with the corresponding Killing vector fields $\{X^a_{(1)}, X^a_{(2)}, X^a_{(3)}\}$. This setting allows us to construct (see technical remarks in \cite{BSJ22,BJS26}) the standard spherical coordinate system $(t,r,\theta,\varphi)$. Furthermore, we assume that the spacetime possesses a ``regular center'', a notion which comprises two logically distinct requirements: (1) the spacetime contains a fixed point of the $SO(3)$ action, so that the radial coordinate takes values $r \in \left[ 0,\infty \right>$ (thus excluding wormholes and similar ``exotic geometries''), and (2) the relevant curvature invariants are bounded, in the sense specified below and in the statements of the theorems. Thus, the spacetime metric takes the following form
\be\label{eq:metric}
\df s^2 = -A(r) \, \df t^2 + \frac{\df r^2}{B(r)} + r^2 \big( \df\theta^2 + \sin^2\theta \, \df\varphi^2 \big) ,
\ee
where $A$ and $B$ are smooth functions which are either strictly positive or strictly negative on some neighborhood of the center $r=0$ (in most of the results that follow, the smoothness assumption about the metric can be weakened to lower-order differentiability). For convenience, we define auxiliary functions
\be
w(r) \defeq \sqrt{A(r)/B(r)} \qqd \gamma(r) \defeq \frac{(A(r)B(r))'}{A(r)} .
\ee
Unlike in the case of the electromagnetic field minimally coupled to gravity \cite{BSJ22,BJS24}, here we cannot in general conclude that $w(r) = 1$.

\smallskip

The main focus of this paper is the question of existence of regular black hole solutions, i.e.~black hole spacetimes without singularities. However, the very notion of a singularity is notoriously elusive \cite{Geroch68,Earman,Wald,ORGSP16,RGetal25,ZG25}, branching into numerous definitions \cite{ES77,ES79}. We shall investigate the boundedness of curvature invariants, since their divergence is often taken as an indication that incomplete causal geodesics, or more generally curves of bounded acceleration, are inextendible. Within the family of polynomial curvature invariants (without covariant derivatives), a prominent role is played by the Kretschmann scalar $K \defeq R_{abcd} R^{abcd}$, due to its physical interpretation in terms of tidal forces and relation to other invariants. Namely, it was recently shown \cite{DMS25} (see also \cite{SKY25}) that the Ricci scalar $R$, the ``Ricci squared'' $S \defeq R_{ab} R^{ab}$ and the Kretschmann scalar $K$ in a static, spherically symmetric spacetime satisfy simple inequalities $R^2 \le 4S \le 6K$. More concretely, for the metric (\ref{eq:metric}) the Kretschmann scalar can be written as a manifestly nonnegative function,
\be\label{eq:K}
K = \frac{4}{r^4}(1-B)^2 + \frac{2}{r^2} \, (B')^2 + \frac{2}{r^2} \, \frac{B^2(A')^2}{A^2} + \left( \frac{AA'B' - B(A'^2 - 2AA'')}{2A^2} \right)^{\!2}
\ee
and the same holds for the relevant differences of curvature invariants,
\begin{align}
4S - R^2 & = \frac{B^2}{A^2} \left( A'' + \frac{2A(1-B)}{r^2 B} - \frac{A'(A'B - AB')}{2AB} \right)^{\!2} + \frac{2(AB' - A'B)^2}{r^2 A^2} , \\
3K - 2S & = \frac{2B^2}{A^2} \left( A'' + \frac{rA'(AB' - A'B) - A(AB' + A'B)}{2rAB} \right)^{\!2} + \nonumber \\
 & \hspace{10pt} + \frac{5B^2}{2r^2 A^2} \left( A' + \frac{A \big( 4(1-B) - 3rB' \big)}{5rB} \right)^{\!2} + \frac{8 \big( 2(1-B) + rB' \big)^2}{5r^4} .
\end{align}
Inequalities between the polynomial curvature invariants were systematically studied and generalized in \cite{ISm26}, where it was shown that all Zakhary--McIntosh invariants in a spherically symmetric spacetime are bounded by suitable powers of the Kretschmann scalar.

\smallskip

In addition to requiring curvature invariants to remain bounded, we shall also impose suitable smoothness conditions (see discussion in \cite{AS26}). First of all, given that the metric components are $C^2$ and the Kretschmann scalar (\ref{eq:K}) is bounded, it follows that
\be
B(r) = 1 + O(r^2) \qqd B'(r) = O(r) \qqd A'(r)/A(r) = O(r) \qqd A''(r)/A(r) = O(1) ,
\ee
as $r \to 0^+$. Secondly, for most of the results the Ricci scalar is required to be $C^n$ with $n$ a small integer (typically 1 or 2), while in certain special subcases the proof relies on Ricci's analyticity (which, presumably, may be weakened by a different argument).

\smallskip

Finally, we assume that the electromagnetic field inherits all spacetime symmetries, $\Lie_K \F = 0$ for all $K^a \in \{k^a, X^a_{(1)}, X^a_{(2)}, X^a_{(3)}\}$. In this setting, it is convenient to introduce the electric 1-form $\pf{E} \defeq -i_k \F$ and the magnetic 1-form $\pf{B} \defeq i_k{\hdg\F}$, with respect to the Killing vector field $k^a$. Symmetry inheritance assumptions are sufficient to deduce (see Appendix \ref{app:Sym}) that $\pf{E} = E_r(r) \, \df r$ and $\pf{B} = B_r(r) \, \df r$. Thus, the electromagnetic 2-form $\F$ and its Hodge dual ${\hdg\F}$ may be decomposed as
\begin{align}
\F & = -E_r(r) \, \df t \w \df r + \frac{r^2}{w(r)} \, B_r(r) \sin\theta \, \df\theta \w \df\varphi , \\
{\hdg\F} & = B_r(r) \, \df t \w \df r + \frac{r^2}{w(r)} \, E_r(r) \, \sin\theta \, \df\theta \w \df\varphi .
\end{align}
The associated electromagnetic invariants are
\be
\FF = \frac{2}{w^2} \, (B_r^2 - E_r^2) \qqd \GG = \frac{4}{w^2} \, E_r B_r .
\ee
Since the electromagnetic field is introduced via $\F = \df\A$, one of the generalized Maxwell equations, $\df\F = 0$, is unaffected by the choice of the nonminimal coupling $\ell$. This immediately implies that $r^2 B_r(r)/w(r)$ is a constant. Moreover, the magnetic charge $P$ may be defined via integral
\be
P \defeq \frac{1}{4\pi} \oint_{\mathcal{S}} \F
\ee
over the 2-sphere $\mathcal{S}$ (orbit of the $SO(3)$ action), so that
\be\label{eq:Br}
B_r(r) = \frac{P}{r^2} \, w(r) .
\ee
The nontrivial part of the electromagnetic field is the function $E_r(r)$, which will be analyzed case by case.

\section{Type A terms} 

\subsection{The \texorpdfstring{$R\FF$}{RF} coupling} 

We begin our analysis with the simplest nonminimal coupling, Lagrangian term $\ell(R,\FF) = \lambda R\FF$, with the coupling constant $\lambda \ne 0$. The corresponding gravitational field equation is
\be\label{eq:EinstA}
G_{ab} + \Lambda g_{ab} = 8\pi (1 - \lambda R) T^{\mathrm{(Max)}}_{ab} + \lambda \big( \nab{a} \nab{b} \FF - g_{ab} \Box\FF - \FF R_{ab} \big) ,
\ee
while the generalized Maxwell's equations are
\be\label{eq:MaxA}
\nab{a} {\hdg F}^{ab} = 0 , \quad \nab{a} \big( (1 - \lambda R) F^{ab} \big) = 0 ,
\ee
which can be written with differential forms as
\be\label{eq:dfMax}
\df\F = 0 , \quad \df{\hdg\big( (1 - \lambda R) \F \big)} = 0 .
\ee
Analogous to the magnetic charge, introduced above, the electric charge $Q$ can be defined with the integral
\be
Q \defeq \frac{1}{4\pi} \oint_{\mathcal{S}} (1 - \lambda R) \, {\hdg\F} ,
\ee
over the 2-sphere $\mathcal{S}$ (orbit of the $SO(3)$ action). Second generalized Maxwell's equation (\ref{eq:dfMax}) implies that $(1-\lambda R(r)) r^2 E_r(r)/w(r)$ is a constant, which leads to the formal expression for the electric field,
\be\label{eq:Er}
E_r(r) = \frac{w(r)}{1 - \lambda R(r)} \, \frac{Q}{r^2} .
\ee
The analysis revolves around the trace of the gravitational field equation, which in this case reads
\be\label{eq:trace}
R - 4\Lambda = \lambda \big( R\FF + 3\Box\FF \big) .
\ee
Note that the D'Alembertian of the electromagnetic invariant $\FF$ may be written as
\be
\Box \FF = \frac{1}{\sqrt{-g}} \, \dd_\mu \big( \sqrt{-g} g^{\mu\nu} \dd_\nu \FF \big) = \frac{(r^2 w B \FF')'}{r^2 w} .
\ee
In the neutral case, when $Q = 0$ and $P = 0$, the electromagnetic field is identically zero and solution of the field equations is again standard (A)dS-Schwarzschild spacetime, which does not have a regular center. We shall treat the charged solutions, starting from the technically simplest magnetically charged black holes, then move to physically most relevant electrically charged black holes and finish with the technically most challenging ones, dyonic black holes. Although all three results could, in principle, be subsumed under a single theorem, we find it considerably more illuminating to discuss the cases separately.

\smallskip

\btm\label{tm:P1}
Suppose that the spacetime is a static, spherically symmetric solution of the field equations (\ref{eq:EinstA}) and (\ref{eq:MaxA}), with the magnetic charge $P \ne 0$ and no electric charge, $Q = 0$. Then the center cannot be regular, in the sense that there is no neighborhood of $r=0$ on which the Kretschmann scalar $K$ is bounded and the Ricci scalar $R$ is a $C^1$ function.
\etm

\noindent
\emph{Proof}. Absence of the electric charge implies $E_r = 0$, while the magnetic field $B_r$ is formally given by the equation (\ref{eq:Br}). This implies that the nontrivial electromagnetic invariant is reduced to
\be
\FF = \frac{2P^2}{r^4} ,
\ee
and the trace equation (\ref{eq:trace}) takes form
\be\label{eq:trmag}
R - 4\Lambda = \frac{2\lambda P^2}{r^4} \left( R + \frac{6}{r^2} \, (6B - r\gamma) \right) ,
\ee
Regularity assumptions imply $R = O(1)$, $B = 1 + O(r^2)$ and $r\gamma = O(r^2)$ as $r \to 0^+$. These, in turn, imply that the right hand side of the equation (\ref{eq:trmag}) is unbounded, while the left hand side is bounded as $r \to 0^+$, which is a contradiction. \qed

\medskip

\btm\label{tm:Q1}
Suppose that the spacetime is a static, spherically symmetric solution of the field equations (\ref{eq:EinstA}) and (\ref{eq:MaxA}), with the electric charge $Q \ne 0$ and no magnetic charge, $P = 0$. Then the center cannot be regular, in the sense that there is no neighborhood of $r=0$ on which all of the following hold: (i) the Kretschmann scalar $K$ is bounded; (ii) the Ricci scalar satisfies $R(r) \ne 1/\lambda$ for all $r > 0$; (iii) $R$ is a $C^2$ function if $\lim_{r\to 0^+} R(r) \ne 1/\lambda$, and $R$ is an analytic function if $\lim_{r\to 0^+} R(r) = 1/\lambda$.
\etm

\noindent
\emph{Proof}. Absence of the magnetic charge implies $B_r = 0$, while the electric field $E_r$ is formally given by the equation (\ref{eq:Er}). This implies that the nontrivial electromagnetic invariant is
\be
\FF = -\frac{2Q^2}{r^4} \, \frac{1}{(1 - \lambda R)^2} ,
\ee
while its D'Alembertian may be conveniently written as
\be\label{eq:BoxFrhosigma}
\Box \FF = \frac{4Q^2}{r^4 (1 - \lambda R)^2} \left( \rho(r) + \sigma(r) + \frac{\gamma(r)}{r} \right) ,
\ee
where
\begin{align}
\rho(r) & = -3B \left( \frac{1}{r^2} + \left( \frac{1}{r} - \frac{\lambda R'}{1 - \lambda R} \right)^{\!2} \right) , \\
\sigma(r) & = -\frac{\lambda}{2(1 - \lambda R)} \, \big( \gamma R' + 2BR'' \big) .
\end{align}
Thus, the trace equation (\ref{eq:trace}) takes form
\be\label{eq:trel}
R - 4\Lambda = \frac{2\lambda Q^2}{r^4 (1-\lambda R)^2} \, \big( {-R} + 6(\rho + \sigma + \gamma/r) \big) . 
\ee
Regularity assumptions imply $R = c_0 + O(r)$ with some constant $c_0$, $B = 1 + O(r^2)$ and $\gamma = O(r)$ as $r \to 0^+$. If $c_0 \ne 1/\lambda$, then $\rho = \Omega(r^{-2})$ and $\sigma = o(\rho)$ as $r \to 0^+$, which implies that the right hand side of the equation (\ref{eq:trel}) is unbounded, while the left hand side is bounded as $r \to 0^+$, which is a contradiction. If $c_0 = 1/\lambda$, then there is a delicate question if some ``accidental canceling'' between the terms in $\rho$ and $\sigma$ may happen. By analyticity of the Ricci scalar, we have the Taylor series with remainder
\be
R(r) = c_0 + c_n r^n + O(r^{n+1}) \ \ \textrm{as} \ \ r \to 0^+ ,
\ee
where $c_0$ and $c_n$ are constants, and $n > 0$ is the smallest positive integer such that $c_n \ne 0$. If there would be no such integer $n$, by analyticity assumption in $(iii)$, the Ricci scalar would be constant and equal to $1/\lambda$ on some neighborhood of $r = 0$, in contradiction with the regularity assumption $(ii)$. Now, as $1 - \lambda R = -\lambda c_n r^n + O(r^{n+1})$, $R' = n c_n r^{n-1} + O(r^n)$ and $R'' = n(n-1)c_n r^{n-2} + O(r^{n-1})$, it follows that $\rho(r) = -3B(1 + (n+1)^2)r^{-2} + O(r^{-1})$ and $\sigma(r) = Bn(n-1) r^{-2} + O(r^{-1})$, which proves that there is no cancellation. Thus, regularity assumptions again imply that the right hand side of the equation (\ref{eq:trel}) is unbounded, while the left hand side is bounded as $r \to 0^+$, which is a contradiction. \qed

\bigskip

\btm\label{tm:QP1}
Suppose that the spacetime is a static, spherically symmetric solution of the field equations (\ref{eq:EinstA}) and (\ref{eq:MaxA}), with the electric charge $Q \ne 0$ and the magnetic charge $P \ne 0$. Furthermore, suppose that Ricci scalar $R$ is a $C^2$ function on some neighborhood of $r=0$ and $c_0 \defeq \lim_{r\to 0^+} R(r)$. Then the center cannot be regular, in the sense that there is no neighborhood of $r=0$ on which all of the following hold: (i) the Kretschmann scalar $K$ is bounded; (ii) the Ricci scalar satisfies $R(r) \ne 1/\lambda$ for all $r > 0$; (iii) $c_0 \ne 1/\lambda$ and $(Q/P)^2 \ne (1 - \lambda c_0)^2$, or $c_0 = 1/\lambda$ and $R$ is an analytic function.
\etm

\noindent
\emph{Proof}. Using notation from the previous proof, electromagnetic invariant $\FF$ and its D'Alem\-ber\-tian may be written as
\begin{align}
\FF & = \frac{2}{r^4} \left( -\frac{Q^2}{(1 - \lambda R)^2} + P^2 \right) , \\
\Box \FF & = \frac{4Q^2}{r^4 (1 - \lambda R)^2} \left( \rho(r) + \sigma(r) + \frac{\gamma(r)}{r} \right) + \frac{4P^2}{r^6} \, \big( 6B - r \gamma(r) \big) .
\end{align}
Suppose first that $c_0 \ne 1/\lambda$ and $(Q/P)^2 \ne (1 - \lambda c_0)^2$. Then the regularity assumptions imply $\FF = O(r^{-4})$ and
\be
\Box\FF = \frac{24 B}{r^6} \left( -\frac{Q^2}{(1 - \lambda R)^2} + P^2 \right) + O(r^{-5}) ,
\ee
which, inserted into the trace equation (\ref{eq:trace}), produce a contradiction. Furthermore, suppose that $c_0 = 1/\lambda$, Ricci scalar is an analytic function on some neighborhood of $r=0$, and
\be
R(r) = c_0 + c_n r^n + O(r^{n+1}) \ \ \textrm{as} \ \ r \to 0^+ ,
\ee
as in the previous proof. Then the dominant terms are
\be
\FF = -\frac{2Q^2}{(\lambda c_n)^2 r^{2(n+2)}} + \dots , \quad \Box\FF = -\frac{4Q^2 B}{(\lambda c_n)^2 r^{2(n+2)}} \, \frac{(n+2)(2n+3)}{r^2} + \dots ,
\ee
which again leads to a contradiction in the trace equation (\ref{eq:trace}).
\qed

\bigskip

The remaining dyonic subcase, not covered by the previous theorem, is the one in which $c_0 \ne 1/\lambda$ and $(Q/P)^2 = (1 - \lambda c_0)^2$. Here we have
\be
\frac{Q^2}{(1-\lambda R)^2} = P^2 + \frac{2\lambda c_n Q^2}{(1 - \lambda c_0)^3} \, r^n + O(r^{n+1}) ,
\ee
so that
\begin{align}
\FF & = -\frac{4\lambda c_n Q^2}{(1 - \lambda c_0)^3} \, r^{n-4} + O(r^{n-3}) , \\
\Box\FF & = -\dfrac{4P^2\lambda B c_n}{(1-\lambda c_0)} \, (n-3)(n-4) r^{n-6} + O(r^{n-5}) .
\end{align}
Although three special cases, $n \in \{1,2,5\}$, seem to lead again to the no-go conclusion, cases $n \in \{3,4\}$ demand further analysis of the subleading terms, while for $n > 6$ (``very flat'' Ricci scalar around center) the trace analysis cease to be effective and one needs to find some other approach. We remark that in the case when $R = c_0$ is constant and charge relation $(Q/P)^2 = (1 - \lambda c_0)^2$ holds, it follows that $\FF = 0$ identically, which leads to a rescaled (A)dS-Reissner--Nordstr\"om solution.

\subsection{Some generalizations} 

Part of the previous analysis may be extended to a slightly generalized nonminimal coupling term of the form $\ell = f(R)\FF$, with some function $f$. The corresponding gravitational field equation is
\be\label{eq:Einst2}
G_{ab} + \Lambda g_{ab} = 8\pi \big( 1 - f(R) \big) T^{\mathrm{(Max)}}_{ab} + \big( \nab{a} \nab{b} - g_{ab} \Box - R_{ab} \big) (\dd_R f(R) \FF) ,
\ee
whose trace gives
\be\label{eq:trace2}
R - 4\Lambda = \dd_R f(R) R \FF + 3\Box\big( \dd_R f(R)\FF \big) ,
\ee
while the generalized Maxwell's equations, written with differential forms, are
\be\label{eq:Max2}
\df\F = 0 , \quad \df{\hdg\big( (1 - f(R)) \F \big)} = 0 .
\ee
The electric charge $Q$ can be defined with the integral
\be
Q \defeq \frac{1}{4\pi} \oint_{\mathcal{S}} \big( 1 - f(R) \big) \, {\hdg\F} .
\ee
Second generalized Maxwell's equation implies that $(1 - f(R))r^2 E_r(r)/w(r)$ is a constant, implying that the electric field can be expressed as
\be
E_r(r) = \frac{w(r)}{1 - f(R)} \, \frac{Q}{r^2} .
\ee
Whereas in general we need to be careful about the smoothness assumptions, e.g.~one should demand function $f$ to be class $C^3$ due to presence of terms like $\Box\,\dd_R f(R)$, in the electric field an additional issue stems from the zeros of the function $1 - f(R)$, that is points at which $R(r) \in f^{-1}(\{1\})$.

\smallskip

The magnetic case admits a straightforward generalization, as stated in the following theorem.

\btm
Let $f : \rr \to \rr$ be a class $C^3$ function and suppose that the spacetime is a static, spherically symmetric solution of the field equations (\ref{eq:Einst2}) and (\ref{eq:Max2}), with the magnetic charge $P \ne 0$ and no electric charge, $Q = 0$. Then the center cannot be regular, in the sense that there is no neighborhood of $r=0$ on which the Kretschmann scalar $K$ is bounded, the Ricci scalar $R$ is a $C^2$ function, and $\lim_{r\to 0^+} \dd_R f(R) \ne 0$.
\etm

\noindent
\emph{Proof}. We have simply $\FF = 2P^2 r^{-4}$, while the dominant contribution to the D'Alembertian term in (\ref{eq:trace2}) is $\Box\big( \dd_R f(R)\FF \big) = 24P^2 B \dd_R f(R) r^{-6} + O(r^{-5})$. Such functions immediately lead to a contradiction in the trace equation (\ref{eq:trace2}) under regularity assumptions stated in the theorem. \qed

\medskip

An example of a magnetically charged regular black hole analyzed in \cite{Sert16} is a solution of a theory with nonminimal coupling $\ell = f(R)\FF$ (with the Lagrangian constructed via reverse engineering), which satisfies $\lim_{r\to 0^+} \dd_R f(R) = 0$, thus evading the assumptions of the no-go theorem above.

\smallskip

The analysis of the electric case becomes far more challenging, as the electromagnetic invariant
\be
\FF = -\frac{2Q^2}{(1 - f(R(r)))^2 r^4}
\ee
harbors the factor $1 - f(R)$ in the denominator. Given that the Ricci scalar $R$ is a $C^2$ function, and $f$ is a $C^3$ function such that $\lim_{r\to 0^+} f(R) \ne 1$ and $\lim_{r\to 0^+} \dd_R f(R) \ne 0$ we may reach inconsistency of regularity assumptions with the trace equation (\ref{eq:trace2}) analogous to the one revealed in the proof of the Theorem \ref{tm:Q1}: the term $\dd_R f(R) R\FF$ is of the order $O(r^{-4})$, while the term $\Box(\dd_R f(R) \FF)$ is of the order $O(r^{-6})$ as $r \to 0^+$. However, the analysis of the special subcases excluded here, is far more intricate and we leave it for future work.

\smallskip

There is another family of Lagrangians which admit relatively straightforward generalization (at least in the electrically charged case), those with nonminimal coupling term of the form $\ell = f(R,\GG)$. The gravitational field equation is
\be\label{eq:Einst3}
G_{ab} + \Lambda g_{ab} = 8\pi T^{\mathrm{(Max)}}_{ab} + \big( \nab{a} \nab{b} - g_{ab} \Box - R_{ab} \big) \dd_R f + \frac{1}{2} \, (f - \GG \, \dd_\GG f) g_{ab}
\ee
while the generalized Maxwell's equations are
\be\label{eq:Max3}
\df\F = 0 \qqd \df({\hdg\F} + \dd_\GG f \, \F) = 0 .
\ee
In analogy with the previous cases, the electric charge $Q$ is defined with the integral
\be
Q \defeq \frac{1}{4\pi} \oint_{\mathcal{S}} ({\hdg\F} + \dd_\GG f \, \F) .
\ee

\btm
Let $f : \rr^2 \to \rr$ be a class $C^3$ function, such that $f(R,0) = 0$ for all $R \in \rr$. Furthermore, suppose that the spacetime is a static, spherically symmetric solution of the field equations (\ref{eq:Einst3}) and (\ref{eq:Max3}), with the electric charge $Q \ne 0$ and no magnetic charge, $P = 0$. Then the solution is (A)dS-Reissner--Nordstr\"om spacetime.
\etm

\noindent
\emph{Proof}. Absence of the magnetic charge, $P = 0$, implies $B_r = 0$ and $\GG = 0$. Hence, we immediately have $\df\F = 0$ and, as $\partial_\GG f$ is a radially dependent function, $\df(\partial_\GG f \F) = 0$. The remaining Maxwell's equation $\df{\hdg\F} = 0$, together with the definition of the electric charge, implies that the electric field is $E_r(r) = w(r) Q r^{-2}$. Finally, assumption $f(R,0) \equiv 0$ reduces gravitational field equation to the Einstein field equation with the Maxwell's electromagnetic energy-momentum tensor. From here, by standard derivation, we get that $w = 1$ and $g_{ab}$ is the (A)dS-Reissner--Nordstr\"om metric. \qed

\medskip

Previous result may be interpreted in the sense that nonminimal coupling $\ell = f(R,\GG)$ becomes ``stealth'' in the purely electric case (for a broader discussion on stealth electromagnetic fields see \cite{ISm18}). Due to the very peculiar nature of such a Lagrangian, in the opposite case with magnetic charge $P \ne 0$ and electric charge $Q = 0$, the magnetic field is given by $B_r(r) = w(r) P r^{-2}$, whereas the electric field $E_r(r) = -(\dd_\GG f) B_r(r)$ is not necessarily zero. We note in passing that this phenomenon, namely a nonvanishing electric field with zero electric charge, may be viewed as the counterpart of the solutions encountered in a family of nonlinear electromagnetic Lagrangians analyzed in \cite{BJS26}, for which the electric charge is nonzero whereas the electric field vanishes. Furthermore, second electromagnetic invariant is $\GG = -4(\dd_\GG f) P^2 r^{-4}$, while the trace of the gravitational field equation reads
\be
R - 4\Lambda = (3\Box + R) \dd_R f + 2(\GG \, \dd_\GG f - f) .
\ee
For example, if the nonminimal coupling is of the form $\ell = h(R) \GG$, then $\GG = -4h(R) P^2 r^{-4}$ which, under mild smoothness assumptions about the function $h$, again leads to a no-go result based on inconsistency of the trace equation, similar to the theorems in subsection 3.1. However, the more general case with $\ell = f(R,\GG)$ becomes far less tractable and we shall not pursue further analysis here.

\section{Type B term} 

We turn to type B term, the nonminimal coupling of the form $\ell = \lambda R_{ab} \tensor{F}{^a_c} F^{bc}$. Here we have gravitational field equation
\begin{align}\label{eq:EinstB}
G_{ab} + \Lambda g_{ab} & = 8\pi T_{ab}^{\mathrm{(Max)}} + \lambda \Big( {-\frac{1}{2}} \, \Box(F_{ap} \tensor{F}{_b^p}) - \frac{1}{2} \, g_{ab} \nab{c} \nab{d} (\tensor{F}{^c_p} F^{dp}) + \nab{c} \nab{(a} (F_{b)p} F^{cp}) + \nonumber \\
 & \hspace{20pt} + \frac{1}{2} \, g_{ab} R_{pq} \tensor{F}{^p_c} F^{qc} - R^{pq} F_{ap} F_{bq} - 2F_{pq} \tensor{R}{_{(a}^p} \tensor{F}{_{b)}^q} \Big)
\end{align}
whose trace may be written as
\be\label{eq:Btr}
R - 4\Lambda = \lambda \Big( \frac{1}{2} \, \Box\FF + \nab{a} \nab{b} (F^{ac} \tensor{F}{^b_c}) + R^{bc} F_{ab} \tensor{F}{^a_c} \Big) ,
\ee
and the generalized Maxwell equations are
\be\label{eq:MaxB}
\nab{a} {\hdg F}^{ab} = 0 , \quad \nab{a} \big( F^{ab} + \lambda \tensor{R}{^{[a}_c} F^{b]c} \big) = 0 .
\ee
Given that we introduce an auxiliary 2-form $Z_{ab} \defeq F_{ab} + \lambda \tensor{R}{_{[a}^c} \tensor{F}{_{b]}_c}$, we can write the latter equation as $\df{\hdg\pf{Z}} = 0$ and define the electric charge with the integral
\be
Q \defeq \frac{1}{4\pi} \oint_{\mathcal{S}} {\hdg\pf{Z}} .
\ee
Furthermore, as
\be
Z_{tr} = \Big( 1 - \frac{\lambda}{2} \, \big( \tensor{R}{_t^t} + \tensor{R}{_r^r} \big) \Big) F_{tr} \qqd Z_{\theta\varphi} = \Big( 1 - \frac{\lambda}{2} \, \big( \tensor{R}{_\theta^\theta} + \tensor{R}{_\varphi^\varphi} \big) \Big) F_{\theta\varphi} ,
\ee
equation $\df{\hdg\pf{Z}} = 0$ implies that the electric field can be expressed as
\be
E_r(r) = \frac{w(r)}{1 - \frac{\lambda}{2} \, \big( \tensor{R}{_t^t} + \tensor{R}{_r^r} \big)} \, \frac{Q}{r^2} .
\ee
This leads us to the constraint results for the type B terms.
\btm
Suppose that the spacetime is a static, spherically symmetric solution of the field equations (\ref{eq:EinstB}) and (\ref{eq:MaxB}), with the magnetic charge $P \ne 0$ and no electric charge, $Q = 0$. Then the center cannot be regular, in the sense that there is no neighborhood of $r=0$ on which the Kretschmann scalar $K$ is bounded and the Ricci scalar $R$ is a $C^1$ function.
\etm

\noindent
\emph{Proof}. Under given assumptions we have $E_r = 0$ and $B_r = Pw(r)/r^2$, so that the trace equation becomes
\be
R - 4\Lambda = \frac{2\lambda P^2}{r^6} \, (1 + 8B(r) - 2r\gamma(r)) ,
\ee
which is inconsistent with the given regularity assumptions. \qed

\medskip

The electric case, with $Q \ne 0$ and $P = 0$ is bit more challenging due to complexity of expressions. If one tries to follow the same approach as in the magnetic case and evaluate the right hand side of (\ref{eq:Btr}) directly in terms of the metric functions, the resulting expression is too involved to isolate the leading divergent term (or perform detailed analysis of all possible subcases). However, as bounded Kretschmann scalar implies that $A(r) = a_0 + a_2 r^2 + O(r^3)$ with $a_0 \ne 0$ and $B(r) = 1 + b_2 r^2 + O(r^3)$ as $r \to 0^+$, a brute force insertion reveals the dominant contribution on the right hand side of the trace equation,
\be
R - 4\lambda = -\frac{18\lambda (a_0 Q)^2}{(a_0 + (2a_2 + a_0 b_2)\lambda)^2} \, r^{-6} + O(r^{-5})
\ee
which again leads to a contradiction and implies that a similar no-go result holds.

\section{Type C term} 

Finally, we look at the type C term, the nonminimal coupling of the form $\ell = \lambda R_{abcd} F^{ab} F^{cd}$. Here we have gravitational field equation
\begin{align}\label{eq:EinstC}
G_{ab} + \Lambda g_{ab} & = 8\pi T_{ab}^{\mathrm{(Max)}} + \lambda \Big( {-3}\tensor{R}{_{(a}^c^d^e} F_{b)c} F_{de} - \nonumber \\
 & \hspace{20pt} - 2 \nab{c} \nab{d} (\tensor{F}{_{(a}^c} \tensor{F}{_{b)}^d}) + \frac{1}{2} \, (R_{cdef} F^{cd} F^{ef}) g_{ab} \Big)
\end{align}
whose trace may be written as
\be
R - 4\Lambda = \lambda \Big( R_{abcd} F^{ab} F^{cd} + 2 \nab{a} \nab{b} (F^{ac} \tensor{F}{^b_c}) ,\Big)
\ee
and the generalized Maxwell equations are
\be\label{eq:MaxC}
\nab{a} {\hdg F}^{ab} = 0 , \quad \nab{a} \big( F^{ab} - \lambda R^{abcd} F_{cd} \big) = 0 .
\ee
Given that we introduce an auxiliary 2-form $Z_{ab} \defeq F_{ab} - \lambda \tensor{R}{_a_b^c^d} \tensor{F}{_c_d}$, we can write the latter equation as $\df{\hdg\pf{Z}} = 0$ and define the electric charge with the integral
\be
Q \defeq \frac{1}{4\pi} \oint_{\mathcal{S}} {\hdg\pf{Z}} .
\ee
Furthermore, as
\be
Z_{tr} = \big( 1 - 2\lambda \tensor{R}{_t_r^t^r} \big) F_{tr} \qqd Z_{\theta\varphi} = \big( 1 - 2\lambda \tensor{R}{_\theta_\varphi^\theta^\varphi} \big) F_{\theta\varphi}
\ee
equation $\df{\hdg\pf{Z}} = 0$ implies that the electric field can be expressed as
\be
E_r(r) = \frac{w(r)}{1 - 2\lambda \tensor{R}{_t_r^t^r}} \, \frac{Q}{r^2} .
\ee
This leads us to the constraint results for the type C terms.

\btm
Suppose that the spacetime is a static, spherically symmetric solution of the field equations (\ref{eq:EinstC}) and (\ref{eq:MaxC}), with the magnetic charge $P \ne 0$ and no electric charge, $Q = 0$. Then the center cannot be regular, in the sense that there is no neighborhood of $r=0$ on which the Kretschmann scalar $K$ is bounded and the Ricci scalar $R$ is a $C^1$ function.
\etm

\noindent
\emph{Proof}. Under given assumptions we have $E_r = 0$ and $B_r = Pw(r)/r^2$, so that the trace equation becomes
\be
R - 4\Lambda = \frac{2\lambda P^2}{r^6} \, (2 + 4B(r) - r\gamma(r))
\ee
which is inconsistent with the given regularity assumptions. \qed

\medskip

Just as with the type B term, the electric case, with $Q \ne 0$ and $P = 0$, is bit more challenging due to complexity of expressions. Bounded Kretschmann scalar implies that $A(r) = a_0 + a_2 r^2 + O(r^3)$ with $a_0 \ne 0$ and $B(r) = 1 + b_2 r^2 + O(r^3)$ as $r \to 0^+$, so a brute force insertion reveals the dominant contribution on the right hand side of the trace equation,
\be
R - 4\lambda = -\frac{12\lambda (a_0 Q)^2}{(a_0 + 2a_2 \lambda)^2} \, r^{-6} + O(r^{-5})
\ee
which again leads to a contradiction and implies that a similar no-go result holds.

\section{Discussion} 

Using the simple ``trace approach", we investigated constraints on regular black holes in EFT-motivated theories with nonminimally coupled gravitational and electromagnetic interactions. Focusing on specific types of nonminimal couplings, we have shown that individually they do not lead to regular black hole solutions. More broadly, taking into account the existing no-go results for NLE theories \cite{Bronnikov00, BSJ22, BJS26}, these results narrow down the possibility of black hole regularisation for certain classes of non-Maxwellian fields. However, even within the studied set of theories, several scenarios are still not covered by the theorems presented in this work. The main open questions can be summarized as follows:

\medskip

1. Fine-tuned dyonic case.

\smallskip

\noindent
Focusing on the $R\FF$ coupling and the dyonic case, Theorem 3.3 leaves a few unresolved cases corresponding to dependent electric and magnetic charges. The relation between the charges is defined with the leading (constant) term in the expansion of the Ricci scalar around the center, which might be related to the regular black hole's mass and charges\footnote{This intuition is supported by the prototypical example of a regular solution, the Bardeen black hole \cite{Bardeen68}.}. This indicates that a mass--charge relation, similar to the one previously observed in the NLE case \cite{BJS26}, may also arise in this framework. Any such relation among the parameters must be supported either by an underlying physical mechanism or a phenomenological justification.

\medskip

2. Further generalizations of A-type couplings.

\smallskip

\noindent
Replacing the linear coupling by an arbitrary function of the Ricci scalar multiplying $\FF$, we showed in Theorem 3.4 that the no-go result for magnetically charged solutions holds only under additional assumptions. An analogous conclusion follows also in the more involved purely electrically charged case. These observations imply that regular black holes may exist once the assumptions of the theorems are relaxed, as illustrated by the solutions in \cite{Sert16}, which violate precisely some of the stated conditions.
More ambitiously, one may attempt to generalize the electromagnetic sector by replacing the Maxwell term with a suitably chosen function of the electromagnetic invariant $\FF$, leading to a nonminimally coupled NLE model.

\medskip

3. Including the $\GG$ electromagnetic invariant.

\smallskip

\noindent
The $\GG$ invariant is itself CP-odd, implying that generic Lagrangians of the form $\ell=f(R, \GG)$ may break the CP symmetry. The constraints derived for A-type terms can be partially extended to this type of couplings, as summarized in Theorem 3.5 and the subsequent discussion. Under the given assumptions, the absence of magnetic charge leads back to the (A)dS Reissner--Nordstr\"om metric. If we set the electric charge to zero, the magnetic charge may give rise to nonzero electric field, making it more difficult to establish a no-go result. Consequently, it holds for a restricted class of theories, leaving room for the construction of nonsingular configurations. A natural, although highly nontrivial, extension would be to allow couplings between the gravitational sector and both electromagnetic invariants.

\medskip

4. Mixing A-, B- and C-type terms.

\smallskip

\noindent
Throughout the paper, we analyzed different classes of couplings separately. However, simultaneous combinations of A-, B- and C-type terms may potentially alter the conclusions of the no-go theorems through cancellations of the ``problematic contributions'' in the trace equation. For the nonminimal coupling
\be
\ell = \lambda_A R\FF + \lambda_B R_{ab} \tensor{F}{^a_c} F^{bc} + \lambda_C R_{abcd} F^{ab} F^{cd}\, ,
\ee
the magnetic case yields the trace equation
\be
R - 4\Lambda = \frac{2P^2}{r^6} \Big( \lambda_B + 2\lambda_C + 4(9\lambda_A + 2\lambda_B + \lambda_C) B + \lambda_A Rr^2 - (6\lambda_A + 2\lambda_B + \lambda_C) r\gamma \Big)
\ee
In order to have consistent equation as $r\to 0^+$, taking into account that $\lim_{r\to 0^+} B(r) = 1$, the necessary condition (though far from sufficient) is that
\be
\lambda_B + 2\lambda_C + 4(9\lambda_A + 2\lambda_B + \lambda_C) = 0
\ee
or equivalently,
\be\label{ABCrel}
12\lambda_A + 3\lambda_B + 2\lambda_C = 0 .
\ee
This immediately excludes Horndeski vector theory, characterized by $\lambda_A=\lambda$, $\lambda_B=-4\lambda$ and $\lambda_C=\lambda$, where $\lambda$ is some common normalization factor. This conclusion is confirmed by the exact solutions \cite{Ver22} and agrees with the result obtained in \cite{CFTS26} via an alternative approach.
On the other hand, there are indications that regular electrically charged configurations in Horndeski theory may exist if Maxwell's electrodynamics is replaced by an appropriate NLE theory \cite{CFTS26}. The regular magnetically charged black hole constructed in \cite{BLZ16} was originally obtained as a solution of the nonminimal Einstein-Yang-Mills theory. However, the solution can be embedded into nonminimally coupled Einstein-Maxwell theory within the class of couplings considered in our analysis. In particular, the theory is equivalent to the linear combination of A-, B-, and C-terms, upon identifying $q_1=\lambda_A$, $2q_2=\lambda_B$ and $q_3=\lambda_C$, where $q_i$ are the coupling constants in the notation of \cite{BLZ16}. The choice of coupling parameters for which regularity is achieved satisfies our condition (\ref{ABCrel}), providing a further independent test of its validity.

\medskip

Overall, the remaining gaps in the theorems leave open two possibilities: either novel regular black hole solutions exist beyond the assumptions considered here, or the trace approach needs to be supplemented by different methods. The former option comes with important caveats. Even if such solutions can be constructed, they might not meet some other physical requirements. Apart from the mentioned mass-charge relations, these black holes may suffer from Ostrogradsky or Laplacian instabilities \cite{Chen_2024, CFTS26}. It is worth emphasizing that the presented no-go results can be easily circumvented by considering more complex types of couplings, which lie outside the scope of our analysis. Some examples include regular magnetically charged black holes found in electromagnetic quasitopological gravities \cite{CM20} and an electrically charged black hole in the model of \cite{CM21_2}. Alternative approaches to singularity resolution with modified gravitational sector, including higher-derivative frameworks such as generalized quasitopological gravity \cite{BCH25, FKSZ25}, as well as quantum gravity inspired models, may also provide a promising route towards regularization\footnote{Singularities could be avoided by either transitioning from Lorentzian to Euclidean metric signature at the black hole horizon \cite{Capozziello:2024ucm} or introducing a regular de Sitter core \cite{Dymnikova:1992ux, Dymnikova:2015yma}.}.

\section*{Acknowledgements}
The research was supported by the European Union --- NextGenerationEU through the National Recovery and Resilience Plan 2021-2026 and the Croatian Science Foundation Project No.~IP-2025-02-8625, \emph{Quantum aspects of gravity}. I.S. acknowledges support from the project``Implementation of cutting-edge research and its application as part of the Scientific Center of Excellence for Quantum and Complex Systems, and Representations of Lie Algebras'', Grant No. PK.1.1.10.0004, co-financed by the European Union through the European Regional Development Fund --- Competitiveness and Cohesion Programme 2021-2027.


\appendix 

\section{Variations}\label{app:Var} 

Here we collect several basic variations of Lagrangian terms, which are essential in derivation of field equations. If $S_{ab}$ is a symmetric tensor, then $S^{ab} \delta g_{ab} = \tensor{S}{^a_c} g^{cb} \delta g_{ab} = -\tensor{S}{^a_c} g_{ab} \delta g^{cb} = -S_{bc} \delta g^{cb}$, which implies an essential ``see-saw rule'' with the indices under variation,  
\be
S^{ab} \delta g_{ab} = -S_{ab} \delta g^{ab} .
\ee
Next, variation of the volume form $\veps$ is
\be
\delta\veps = -\frac{1}{2} \, g_{ab} \, \delta g^{ab} \, \veps .
\ee
Variations of the Christoffel symbol, Riemann tensor, the Ricci tensor and the Ricci scalar may be written as
\begin{align}
\delta\Gamma^a_{bc} & = \frac{1}{2} \, g^{ad} \big( \nab{b} \, \delta g_{cd} + \nab{c} \, \delta g_{bd} - \nab{d} \, \delta g_{bc} \big) , \\
\delta \tensor{R}{^a_b_c_d} & = \nab{c} \, \delta\Gamma^a_{bd} - \nab{d} \, \delta\Gamma^a_{bc} , \\
\delta R_{ab} & = \frac{1}{2} \, \Big( {-g}_{ac} \nab{d} \nab{b} \, \delta g^{cd} - g_{bc} \nab{d} \nab{a} \, \delta g^{cd} - \Box \, \delta g_{ab} + g_{cd} \nab{a} \nab{b} \, \delta g^{cd} \Big) , \\
\delta R & = R_{ab} \, \delta g^{ab} + \nabla_a v^a \qqd v^a = -\nab{b} \, \delta g^{ab} + g_{bc} \nabla^a \delta g^{bc} .
\end{align}
Variations of the Einstein--Hilbert and the Maxwell action are, respectively,
\begin{align}
\delta((R - 2\Lambda) \veps) & = (G_{ab} + \Lambda g_{ab}) \delta g^{ab} \veps + \nabla^a \Theta_a \veps , \\
\delta(\FF\veps) & = 8\pi T^{\mathrm{(Max)}}_{ab} \delta g^{ab} \veps + 4\big( \nab{a}(F^{ab} \delta A_b) - (\nab{a} F^{ab}) \delta A_b \big) \veps ,
\end{align}
where $\Theta_a$ is a boundary term (its details are not relevant for this paper). Next, note that the term $\GG \veps = -2 \, \F \w \F$ is metric-independent, so that
\be
\delta(\GG\veps) = 4\big( {-(\nab{a}{\hdg F}^{ab})} \delta A_b + \nab{a}({\hdg F}^{ab} \delta A_b)  \big) \veps .
\ee
Hence,
\begin{align}
R\delta(\GG\veps) & = 4\big( \nab{a}(R\,{\hdg F}^{ab} \delta A_b) - \nab{a}(R\,{\hdg F}^{ab}) \delta A_b \big) \veps , \\
(\delta R) \GG \veps & = \GG R_{ab} \delta g^{ab} \veps + \nabla^c(\GG v_c) \veps - (v_c \nabla^c \GG) \veps , \\
v_c \nabla^c \GG & = (\nab{a} \nab{b} \GG - g_{ab} \Box\GG) \delta g^{ab} + \nabla^a \widetilde{\Theta}_a ,
\end{align}
with the boundary term $\widetilde{\Theta}_a$.

\section{Symmetries of the electric and the magnetic 1-forms}\label{app:Sym} 

The spacetime in focus is static, with the corresponding Killing vector field $k^a$, and spherically symmetric, with the corresponding Killing vector fields $\{X^a_{(1)}, X^a_{(2)}, X^a_{(3)}\}$. In the standard spherical coordinate system $(t,r,\theta,\varphi)$, these vector fields may be written as
\begin{align}
k^a & = \dd_t^a , \\
X^a_{(1)} & = -\sin\varphi \, \dd_\theta^a - \cot\theta \cos\varphi \, \dd_\varphi^a , \\
X^a_{(2)} & = \cos\varphi \, \dd_\theta^a - \cot\theta \sin\varphi \, \dd_\varphi^a , \\
X^a_{(3)} & = \dd_\varphi^a .
\end{align}
For completeness, in this appendix we look more closely into the properties of the electric 1-form $\pf{E} = -i_k \F$ and the magnetic 1-form $\pf{B} = i_k {\hdg\F}$. The claim is that given symmetry inheritance assumptions, $\Lie_K \F = 0$ for all four Killing vector fields $K^a$, imply that $\pf{E} = E_r(r) \, \df r$ and $\pf{B} = B_r(r) \, \df r$.

\smallskip

There are two approaches to this problem: (1) a coordinate independent, which relies on some quasi-local properties of the spacetime (presence of the axis of symmetry) and convenient properties of the field equations; (2) a coordinate based, which is essentially pointwise (apart from the initial introduction of local coordinate chart) and is independent of the field equations. The former was used in the classical results on black hole uniqueness theorems \cite{Heusler} and generalized in context of nonlinear electromagnetism \cite{BJS24}. However, when field equations become more tangled, as in the case of nonminimal coupling, the latter option becomes a pragmatic choice, at expanse of ``less elegant'' utilization of the coordinates. Let us look more closely into the both approaches.

\smallskip

(1) The generalized source-free Maxwell equations take form $\df\F = 0$ and $\df{\hdg\Z} = 0$, with the 2-form $Z_{ab} \defeq \dd\LL/\dd F^{ab}$. Assuming that $\LL$ is constructed from symmetry inheriting tensors, assumption $\Lie_K \F = 0$ implies that $\Lie_K \Z = 0$ as well for any Killing vector field $K^a$. We first introduce another 1-form $\pf{H} \defeq i_k {\hdg\Z}$. Then the field equations assure that both $\pf{E}$ and $\pf{H}$ are closed 1-forms, $\df\pf{E} = -\df i_k \F = (-\Lie_k + i_k \df) \F = 0$ and $\df\pf{H} = \df i_k {\hdg\Z} = (\Lie_k - i_k \df) {\hdg\Z} = 0$, thus by Poincar\'e lemma we know that at least locally there are scalars $\Phi$ and $\Psi$, such that $\pf{E} = -\df\Phi$ and $\pf{H} = -\df\Psi$. Next, if we apply the identity
\be
\df i_X i_Y = i_X i_Y \df + i_{[X,Y]} - i_X \Lie_Y + i_Y \Lie_X
\ee
on 2-forms $\F$ and ${\hdg\Z}$, with $X^a = X^a_{(i)}$ and $Y^a = k^a$, we deduce that $i_{X_{(i)}} i_k \F$ and $i_{X_{(i)}} i_k {\hdg\Z}$ are constant for all $i$. Furthermore, they are zero
on each connected component of spacetime $M$ whose boundary intersects zeros of the corresponding Killing vector field $X^a_{(i)}$ (the axis of symmetry). Hence, it follows that $\Lie_K \Phi = i_K i_k \F = 0$ and $\Lie_K \Psi = i_K i_k {\hdg\Z} = 0$, that is $\Phi = \Phi(r)$ and $\Psi = \Psi(r)$, so that $\pf{E} = -\Phi'(r) \, \df r$ and $\pf{H} = -\Psi'(r) \, \df r$. The tricky part is to deduce the claim for the magnetic 1-form $\pf{B}$. Namely, this is immediate for the classic Maxwell's case, in which $\Z$ is proportional to $\F$, and straightforward for the nonlinear electromagnetism, in which $\Z$ is a linear combination of $\F$ and ${\hdg\F}$. In the case of nonminimal coupling, we have $Z_{ab} = F_{ab} + \tensor{\chi}{_a_b^c^d} F_{cd}$, with the ``susceptibility tensor'' $\chi_{abcd}$ (see discussion in \cite{BL05}). The conclusion subtly depends on the algebraic properties of the tensor $\chi_{abcd}$, which we shall not investigate here. 

\smallskip

(2) Let $\bm{\alpha}$ be 1-form which satisfies $\Lie_K \bm{\alpha} = 0$ for all $K^a \in \{k^a, X^a_{(1)}, X^a_{(2)}, X^a_{(3)} \}$. First of all, $\Lie_k \bm{\alpha} = 0$ and $\Lie_{X_{(3)}} \bm{\alpha} = 0$ imply that
\be
\bm{\alpha} = \alpha_t(r,\theta) \, \df t + \alpha_r(r,\theta) \, \df r + \alpha_\theta(r,\theta) \, \df\theta + \alpha_\varphi(r,\theta) \, \df\varphi .
\ee
Then, using the identity
\be\label{id:iYLX}
i_Y \Lie_X \bm{\alpha} = \Lie_X i_Y \bm{\alpha} - i_{[X,Y]} \bm{\alpha} 
\ee
for $(X^a,Y^a) = (X^a_{(3)},X^a_{(2)})$ we get
\be
0 = \sin\varphi \, \alpha_\theta + \cot\theta \cos\varphi \, \alpha_\varphi ,
\ee
from where it follows that $\alpha_\theta = 0$ and $\alpha_\varphi = 0$. Finally, application of the identity (\ref{id:iYLX}) with $(X^a,Y^a) = (X^a_{(1)},k^a)$ and $(X^a,Y^a) = (X^a_{(1)},\dd_r^a)$ implies that $\dd_\theta \alpha_t = 0$ and $\dd_\theta \alpha_r = 0$. Thus, the most general form of the given 1-form is
\be
\bm{\alpha} = \alpha_t(r) \, \df t + \alpha_r(r) \, \df r .
\ee
In the special case of the electric and the magnetic 1-forms we have additional constraints $i_k \pf{E} = 0$ and $i_k \pf{B} = 0$, which imply $E_t = 0$ and $B_t = 0$, so that $\pf{E} = E_r(r) \, \df r$ and $\pf{B} = B_r(r) \, \df r$, as claimed.

\bibliographystyle{amsalpha}
\bibliography{crnem}

\newcommand{\etalchar}[1]{$^{#1}$}
\providecommand{\bysame}{\leavevmode\hbox to3em{\hrulefill}\thinspace}
\providecommand{\MR}{\relax\ifhmode\unskip\space\fi MR }
\providecommand{\MRhref}[2]{%
  \href{http://www.ams.org/mathscinet-getitem?mr=#1}{#2}
}
\providecommand{\href}[2]{#2}
\begin{thebibliography}{LJMORG25}

\bibitem[ABG00]{ABG00}
E.~Ay{\'o}n-Beato and A.~Garc{\'i}a, \emph{{The Bardeen model as a nonlinear
  magnetic monopole}}, Phys. Lett. \textbf{{\bf B493}} (2000), 149--152.

\bibitem[AS26]{AS26}
T.~Antonelli and M.~Sebastianutti, \emph{{Singularity and differentiability at
  the origin of static and spherically symmetric black holes}}, Phys. Rev. D
  \textbf{{\bf 113}} (2026), no.~6, 064007.

\bibitem[Bar68]{Bardeen68}
J.~M. Bardeen, \emph{{Non-singular General Relativistic Gravitational
  Collapse}}, {Proceeding of the International Conference GR5}, Tbilisi
  University Press, Tbilisi, 1968, p.~174.

\bibitem[BBL08]{BBL08}
A.~B. Balakin, V.~V. Bochkarev, and J.~P.~S. Lemos, \emph{Nonminimal coupling
  for the gravitational and electromagnetic fields: Black hole solutions and
  solitons}, Phys. Rev. D \textbf{77} (2008), 084013.

\bibitem[BCH25]{BCH25}
P.~Bueno, P.~A. Cano, and R.~A. Hennigar, \emph{{Regular black holes from pure
  gravity}}, Phys. Lett. B \textbf{{\bf 861}} (2025), 139260.

\bibitem[BI34]{BI34}
M.~Born and L.~Infeld, \emph{{Foundations of the New Field Theory}}, Proc. R.
  Soc. A \textbf{{\bf 144}} (1934), 425--451.

\bibitem[BJS24]{BJS24}
A.~Bokuli{\'c}, T.~Juri{\'c}, and I.~Smoli{\'c}, \emph{{Hexadecapole at the
  heart of nonlinear electromagnetic fields}}, Class. Quantum Grav.
  \textbf{{\bf 41}} (2024), no.~15, 157002.

\bibitem[BJS26]{BJS26}
\bysame, \emph{{Conundrum of regular black holes with nonlinear electromagnetic
  fields}}, Phys. Rev. D \textbf{{\bf 113}} (2026), no.~2, 024044.

\bibitem[BL05]{BL05}
A.~B. Balakin and J.~P.~S. Lemos, \emph{{Non-minimal coupling for the
  gravitational and electromagnetic fields: A General system of equations}},
  Class. Quantum Grav. \textbf{{\bf 22}} (2005), 1867--1880.

\bibitem[BLZ16]{BLZ16}
Alexander~B. Balakin, Jos\'e P.~S. Lemos, and Alexei~E. Zayats, \emph{Regular
  nonminimal magnetic black holes in spacetimes with a cosmological constant},
  Phys. Rev. D \textbf{93} (2016), 024008.

\bibitem[Bor34]{Born34}
M.~Born, \emph{{On the Quantum Theory of the Electromagnetic Field}}, Proc. R.
  Soc. A \textbf{{\bf 143}} (1934), 410--437.

\bibitem[Bro01]{Bronnikov00}
K.~A. Bronnikov, \emph{{Regular magnetic black holes and monopoles from
  nonlinear electrodynamics}}, Phys. Rev. D \textbf{{\bf 63}} (2001), 044005.

\bibitem[BSJ22]{BSJ22}
A.~Bokuli{\'c}, I.~Smoli{\'c}, and T.~Juri{\'c}, \emph{{Constraints on
  singularity resolution by nonlinear electrodynamics}}, Phys. Rev. D
  \textbf{{\bf 106}} (2022), no.~6, 064020.

\bibitem[Buc79]{Buch79}
H.~A. Buchdahl, \emph{{On a Lagrangian for non-minimally coupled gravitational
  and electromagnetic fields}}, J. Phys. A \textbf{{\bf 12}} (1979),
  1037--1043.

\bibitem[BV14]{BALART201414}
L.~Balart and E.~C. Vagenas, \emph{Regular black hole metrics and the weak
  energy condition}, Physics Letters B \textbf{730} (2014), 14--17.

\bibitem[BZ15]{BZ15}
A.~B. Balakin and A.~E. Zayats, \emph{{Nonminimal black holes with regular
  electric field}}, Int. J. Mod. Phys. D \textbf{{\bf 24}} (2015), no.~09,
  1542009.

\bibitem[CDBB24]{Capozziello:2024ucm}
Salvatore Capozziello, Silvia De~Bianchi, and Emmanuele Battista,
  \emph{{Avoiding singularities in Lorentzian-Euclidean black holes: The role
  of~atemporality}}, Phys. Rev. D \textbf{109} (2024), no.~10, 104060.

\bibitem[CDFTS26]{CFTS26}
C.-Y. Chen, A.~De~Felice, S.~Tsujikawa, and T.~Sano, \emph{Vector horndeski
  black holes in nonlinear electrodynamics}, Phys. Rev. D \textbf{113} (2026),
  024027.

\bibitem[CFT24]{Chen_2024}
Che-Yu Chen, Antonio~De Felice, and Shinji Tsujikawa, \emph{Linear stability of
  vector horndeski black holes}, Journal of Cosmology and Astroparticle Physics
  \textbf{2024} (2024), no.~07, 022.

\bibitem[CM20]{CM20}
Pablo~A. Cano and {\'A}ngel Murcia, \emph{{Electromagnetic Quasitopological
  Gravities}}, JHEP \textbf{10} (2020), 125.

\bibitem[CM21a]{CM21_3}
P.~A. Cano and {\'A}.~Murcia, \emph{{Duality-invariant extensions of
  Einstein--Maxwell theory}}, JHEP \textbf{{\bf 08}} (2021), 042.

\bibitem[CM21b]{CM21}
\bysame, \emph{{Exact electromagnetic duality with nonminimal couplings}},
  Phys. Rev. D \textbf{{\bf 104}} (2021), no.~10, L101501.

\bibitem[CM21c]{CM21_2}
\bysame, \emph{{Resolution of Reissner--Nordstr{\"o}m singularities by
  higher-derivative corrections}}, Class. Quantum Grav. \textbf{{\bf 38}}
  (2021), no.~7, 075014.

\bibitem[CRDEF25]{CRDEF25}
R.~Carballo-Rubio, H.~Delaporte, A.~Eichhorn, and P.~G.~S. Fernandes,
  \emph{Nonminimal light-curvature couplings and black-hole imaging}, Phys.
  Rev. D \textbf{112} (2025), 103016.

\bibitem[DH80]{DH80}
I.~T. Drummond and S.~J. Hathrell, \emph{{QED Vacuum Polarization in a
  Background Gravitational Field and Its Effect on the Velocity of Photons}},
  Phys. Rev. D \textbf{{\bf 22}} (1980), 343.

\bibitem[DK15]{Dymnikova:2015yma}
Irina Dymnikova and Maxim Khlopov, \emph{{Regular black hole remnants and
  graviatoms with de Sitter interior as heavy dark matter candidates probing
  inhomogeneity of early universe}}, Int. J. Mod. Phys. D \textbf{24} (2015),
  no.~13, 1545002.

\bibitem[DMS25]{DMS25}
J.~Draga{\v{s}}evi{\'c}, I.~Moslavac, and I.~Smoli{\'c}, \emph{{Weighing the
  curvature invariants}}, Eur. Phys. J. C \textbf{{\bf 85}} (2025), no.~7, 818.

\bibitem[Dym92]{Dymnikova:1992ux}
I.~Dymnikova, \emph{{Vacuum nonsingular black hole}}, Gen. Rel. Grav.
  \textbf{24} (1992), 235--242.

\bibitem[Ear95]{Earman}
J.~Earman, \emph{{Bangs, Crunches, Whimpers, And Shrieks: Singularities and
  Acausalities in Relativistic Spacetimes}}, Oxford University Press, Oxford,
  1995.

\bibitem[ES77]{ES77}
G.~F.~R. Ellis and B.~G. Schmidt, \emph{{Singular Space-Times}}, Gen. Rel.
  Grav. \textbf{{\bf 8}} (1977), 915--953.

\bibitem[ES79]{ES79}
\bysame, \emph{{Classification of singular space-times}}, Gen. Rel. Grav.
  \textbf{{\bf 10}} (1979), 989--997.

\bibitem[FKSZ25]{FKSZ25}
V.~P. Frolov, A.~Koek, J.~P. Soto, and A.~Zelnikov, \emph{{Regular black holes
  inspired by quasitopological gravity}}, Phys. Rev. D \textbf{{\bf 111}}
  (2025), no.~4, 044034.

\bibitem[Ger68]{Geroch68}
R.~P. Geroch, \emph{{What is a Singularity in General Relativity?}}, Annals
  Phys. \textbf{{\bf 48}} (1968), 526--540.

\bibitem[HE36]{HE36}
W.~Heisenberg and H.~Euler, \emph{{Folgerungen aus der Diracschen Theorie des
  Positrons}}, Z. Phys. \textbf{{\bf 98}} (1936), 714--732.

\bibitem[Heu96]{Heusler}
M.~Heusler, \emph{{Black Hole Uniqueness Theorems}}, Cambridge University
  Press, Cambridge New York, 1996.

\bibitem[Hor76]{HOR76}
G.~W. Horndeski, \emph{Conservation of charge and the einstein–maxwell field
  equations}, J. Math. Phys. \textbf{17} (1976), no.~11, 1980--1987.

\bibitem[LJMORG25]{RGetal25}
H.~C.~D. Lima~Junior, R.~B. Magalh{\~a}es, G.~J. Olmo, and D.~Rubiera-Garcia,
  \emph{{On the resolution of space-time singularities in spherically symmetric
  black holes: geodesic completeness, curvature scalars, and tidal forces}},
  Class. Quantum Grav. \textbf{{\bf 42}} (2025), no.~22, 225004.

\bibitem[MH88]{MMHS88_1}
F.~Müller-Hoissen, \emph{Non-minimal coupling from dimensional reduction of
  the gauss-bonnet action}, Physics Letters B \textbf{{\bf 201}} (1988), no.~3,
  325--327.

\bibitem[MHS88]{MHS88}
F.~M{\"u}ller-Hoissen and R.~Sippel, \emph{{Spherically symmetric solutions of
  the non-minimally coupled Einstein-Maxwell equations}}, Class. Quantum Grav.
  \textbf{{\bf 5}} (1988), 1473--1488.

\bibitem[ORGSP16]{ORGSP16}
G.~J. Olmo, D.~Rubiera-Garcia, and A.~Sanchez-Puente, \emph{{Impact of
  curvature divergences on physical observers in a wormhole
  space{\textendash}time with horizons}}, Class. Quantum Grav. \textbf{{\bf
  33}} (2016), no.~11, 115007.

\bibitem[PS19]{PS19}
P.~Pavlovi{\'c} and M.~Sossich, \emph{{Effect of vacuum polarization on the
  magnetic fields around a Schwarzschild black hole}}, Phys. Rev. D
  \textbf{{\bf 99}} (2019), no.~2, 024011.

\bibitem[RPS23]{RPS23}
K.~Ravi, P.~Pavlovi{\'c}, and A.~Saveliev, \emph{{On the non-minimal coupling
  of magnetic fields with gravity in Schwarzschild spacetime}}, Class. Quantum
  Grav. \textbf{{\bf 40}} (2023), no.~7, 075016.

\bibitem[RT24]{Russo1}
J.~G. Russo and P.~K. Townsend, \emph{{Causality and energy conditions in
  nonlinear electrodynamics}}, JHEP \textbf{{\bf 06}} (2024), 191.

\bibitem[RT26]{Russo2}
\bysame, \emph{{Black holes and causal nonlinear electrodynamics}}, [arXiv:
  2601.07789], 1 2026.

\bibitem[Ser16]{Sert16}
{\"O}.~Sert, \emph{Regular black hole solutions of the non-minimally coupled
  {$Y(R) F^2$} gravity}, J. Math. Phys. \textbf{{\bf 57}} (2016), no.~3,
  032501.

\bibitem[SKN26]{SKY25}
S.~J. Szybka, Y.~Kravetska, and K.~Nikiel, \emph{{Some inequalities among
  curvature invariants}}, Eur. Phys. J. C \textbf{{\bf 86}} (2026), no.~3, 205.

\bibitem[Smo18]{ISm18}
I.~Smoli{\'c}, \emph{{Spacetimes dressed with stealth electromagnetic fields}},
  Phys. Rev. D \textbf{{\bf 97}} (2018), no.~8, 084041.

\bibitem[Smo26]{ISm26}
\bysame, \emph{{Partial Orderings of Curvature Invariants}}, [arXiv:
  2603.08775], 3 2026.

\bibitem[TW88]{TW88}
M.~S. Turner and L.~M. Widrow, \emph{Inflation-produced, large-scale magnetic
  fields}, Phys. Rev. D \textbf{37} (1988), 2743--2754.

\bibitem[V{\etalchar{+}}23]{Vagnozzi:2022moj}
Sunny Vagnozzi et~al., \emph{{Horizon-scale tests of gravity theories and
  fundamental physics from the Event Horizon Telescope image of Sagittarius
  A}}, Class. Quant. Grav. \textbf{40} (2023), no.~16, 165007.

\bibitem[Ver22]{Ver22}
Y.~Verbin, \emph{{Magnetic and electric black holes in the vector-tensor
  Horndeski theory}}, Phys. Rev. D \textbf{106} (2022), no.~2, 024057.

\bibitem[Wal84]{Wald}
R.~Wald, \emph{{General Relativity}}, University of Chicago Press, Chicago,
  1984.

\bibitem[YGZ26]{YGZ26}
Z.~Yin, C.~Gao, and Y.-L. Zhang, \emph{{Photon rings and shadows of black holes
  with non-minimal couplings between curvature and electromagnetic field}},
  [arXiv: 2604.16551], 4 2026.

\bibitem[ZG25]{ZG25}
X.~Zhang and S.~Gao, \emph{{Geodesic completeness, curvature singularities and
  infinite tidal forces}}, [arXiv:2507.04616], 7 2025.

\end{thebibliography}

\end{document}